# Tuning emission energy and fine structure splitting in quantum dots emitting in the telecom O-band

B. Höfer,[1,a] F. Olbrich,[2,a] J. Kettler,[2] M. Paul,[2] J. Höschele,[2] M. Jetter,[2] S. L. Portalupi,[2] F. Ding,[1,3,b] P. Michler[2,c] and O. G. Schmidt,[1,4]

[1]*Institute for Integrative Nanoscience, IFW Dresden, 01069 Dresden, Germany*

[2] *Institut für Halbleiteroptik und Funktionelle Grenzflächen, Center for Integrated Quantum Science and Technology (IQ$^{ST}$) and SCoPE, University of Stuttgart, Allmandring 3, 70569 Stuttgart, Germany*

[3] *Institut für Festkörperphysik, Leibniz Universität Hannover, Appelstrasse 2, 30167 Hannover, Germany*

[4] *Material Systems for Nanoelectronics, TU Chemnitz, 09107 Chemnitz, Germany*

We report on optical investigations of MOVPE-grown InGaAs/GaAs quantum dots emitting at the telecom O-band that were integrated onto uniaxial piezoelectric actuators. This promising technique, which does not degrade the optical quality or performances of the quantum emitters, enables us to tune the quantum dot emission wavelengths and their fine-structure splitting. By spectrally analyzing the emitted light with respect to its polarization, we are able to demonstrate the cancelation of the fine structure splitting within the experimental resolution limit. This work represents an important step towards the high-yield generation of entangled photon pairs at telecommunication wavelength, together with the capability to precisely tune the emission to target wavelengths.

a) B. Höfer and F. Olbrich contributed equally to this work

b) Corresponding Author: F. Ding (f.ding@ifw-dresden.de)

c) Corresponding Author: P. Michler (p.michler@ihfg.uni-stuttgart.de)

Self-assembled semiconductor quantum dots (QDs) are the most promising candidates as sources of on-demand polarization entangled photon pairs, which are highly desired for next generation quantum information and telecommunication applications, e.g. quantum relays and repeaters.[1-3] Furthermore, this technology allows for straightforward on-chip integration[4-6] enabling rapid transfer from proof-of-concept devices to the applied system level. State-of-the-art technology for QDs is nowadays set by GaAs-based structures [7-10] which naturally emit in the NIR wavelength range [10]. In order to transmit single photon-encoded information over long distances and with limited pulse distortion, flying Qbits are expected to emit in the so-called telecommunication O- and C-bands (~1310 nm and ~1550 nm, respectively)[11]. InP-based structures reach emission wavelengths in the telecom range, hence benefitting from low loss fiber communication,[11] but suffering from a lack of effective distributed Bragg reflectors (DBRs). For this reason, efforts have been made in order to extend the GaAs-based technology up to the telecom regime, in order to transfer the leading technology from NIR to a spectral range suitable for both fiber communication and integration with silicon photonics. To this end, single-photon emission in the telecom bands has recently been demonstrated for GaAs-based devices [12,13], as well as the resonant excitation scheme [14] and the creation of entangled photon pairs [15,16]. The strongest limitation in using a QD as source for entangled photons is given by the broken symmetry of as-grown QDs, caused by the anisotropy of strain, composition and shape, which leads to the exciton emission split into two bright excitonic sates. These two states are orthogonally polarized in the linear basis, and their energy difference is generally referred to as the fine-structure splitting (FSS).[17-20] A finite FSS results in an additional phase term in the two-photon polarization state created by the biexciton-exciton radiative cascade. For time-integrated measurements, this effect has to be compensated.[21] To create the entangled states $|\Psi^+\rangle = 1/\sqrt{2}(|H_{XX}H_X\rangle + |V_{XX}V_X\rangle)$, with H and V being the horizontal and vertical polarizations, it is preferable that the FSS is smaller than the radiative lifetime limited linewidth of ~ 1 μeV[22].

In the past few years, several post-growth tuning approaches were developed, such as uniaxial strain induced by piezoelectric materials,[22-24] electric field induced quantum confined Stark effect,[25,26] magnetic field induced Zeeman shifts[27,28] or laser annealing techniques, [29,30] to be able to individually engineer the FSS in semiconductor QDs emitting at $\lambda < 1$ μm. These techniques ultimately lead to a high yield of QDs capable of emitting polarization-entangled photon pairs effectively. Post-growth tuning of the FSS in the telecommunication range has also been demonstrated,[31] the full cancellation of the FSS, however, was not achieved in spectral ranges beyond $\lambda = 1$ μm, yet.

Another advantage of such post-growth tuning techniques is the simultaneous tuning of the FSS and the emission energy, which can be decoupled by adding more than one tuning knob. Such a flexibility in emission energy is beneficial in order to increase the yield of applicable QDs, if a distinct resonance is required, e.g. in remote QD indistinguishability experiments [32] or for implementing hybrid quantum systems [33].

Here, by means of uniaxial strain tuning, we demonstrate the successful elimination of the FSS for telecom-wavelength QDs. The investigated QDs emit at the telecom O-band (1260 – 1360 nm) for which the dispersion in standard silica optical fibers is minimal and the absorption undergoes a local minimum. The sample was fabricated by metal-organic vapor-phase epitaxy (MOVPE) in a commercial AIX 200 laminar flow reactor at a pressure of 100 mbar on a (100) GaAs substrate. Sample fabrication via MOVPE further assists the perspective of a good industrial scalability; in addition to that, the capability of growing on GaAs instead of InP substrates provides the opportunity to increase the source complexity by straight forward integration of binary Bragg mirror systems, which further support the possibility to fabricate high-quality photonic cavity devices as micropillar cavities.[34] As precursors we used TMGa, TMIn, TMAl, and $AsH_3$. After the removal of the oxide at 710°C, we deposited 50 nm of GaAs to ensure a high-quality epitaxial growth surface. This buffer is followed by an $Al_{0.75}Ga_{0.25}As$ sacrificial layer with a thickness of 100 nm that allows the removal of the substrate in a post-growth processing step. The QDs are embedded in a GaAs membrane with an overall thickness of 460 nm. After the deposition of the first GaAs layer, the temperature is lowered from 710°C to 530°C and InGaAs with a nominally equal concentration of Ga and In in the gas phase is introduced for the formation of the QDs. The QDs are then capped by a strain reducing layer of $In_{0.16}Ga_{0.84}As$ to achieve the desired red shift to the telecom O-band. Subsequently, the membrane is completed after the deposition of a GaAs top layer, which eliminates the non-radiative decay channels caused by surface effects. The complete layer stack is shown in Figure 1a. Further details of the QD growth can be found in *Ref.*[12], and information about its structure and morphology in *Ref.*[35] Figure 1b displays a broad-range spectrum of the as-grown sample before the integration onto the piezoelectric substrate. At short wavelengths the wetting layer (WL) emission is observed, while sharp emission lines originating from the QDs are found in the telecom O-band. The micro-photoluminescence (µ-PL) spectroscopy together with a low spatial QD density is sufficient to isolate the emission of single QDs. The observed spectral lines show mainly excitonic behavior, i.e. finite FSSs and linear power dependencies, and the corresponding QD asymmetry tends to align along the [110] crystal axes as observed for similar QD architectures.[36]

To achieve the desirable tunability of the emission energy and FSS the as grown sample (Fig. 1) was further processed, first via vertical wet chemical etching, to define lateral sheet structures of 120 x 160 µm² to be transferred. As a second step the sacrificial layer was removed to create self-standing nanomembranes, which were then integrated onto a PMN-PT ($[Pb(Mg_{1/3}Nb_{2/3})O_3]_{0.72}$-$[PbTiO_3]_{0.28}$) piezoelectric actuator with a combination of thermo-compression bonding and a flip-chip transfer method.[37] Similar structures have previously been used to achieve independent energy and FSS tuning [38] by employing uniaxial or biaxial stress. Here we used uniaxial stress as this has a stronger impact on the FSS. The final device structure is shown in Figure 2a.

For the optical characterization, the samples were mounted in a helium flow-cryostat operating at 4K and were optically excited above the GaAs band gap using a Helium-Neon continuous-wave laser. A confocal microscopy setup equipped with a near infrared objective (numerical aperture of 0.6) was used to collect the emission from single QDs. The QD light was analyzed by a standard 0.5 m spectrometer equipped with a nitrogen cooled InGaAs-CCD array suitable for telecom wavelengths. By inserting a half-wave plate (HWP) and a linear polarizer after the collection lens, polarization-resolved measurements were performed to estimate the FSS.[24] The brightness of typical QD emission lines on the final device structure appears comparable to that before the nanomembrane processing in Figure 1b, which is a clear indication that the QD signal was not influenced by the fabrication steps. To tune the QD emission, variable uniaxial tensile/compressive strain fields can be applied to the QD-containing nanomembranes by changing the voltage $V_P$ across the 300-μm-thick PMN-PT substrate over a range of -200 V < $V_P$ < 1100 V. Note that a tuning voltage $V_P$ in the range of only a few volts can be realized by thin PMN-PT films in an on-chip integrated platform. [6,39]

Figure 2b shows a typical tuning result, where the piezo-voltage $V_P$ is varied from -200 V to 600 V and the QD emission lines shift by about 2 nm, which corresponds to a precise energy tuning rate of 2.5 pm/V. The results of two exemplary polarization resolved photoluminescence measurements are shown in Fig. 2 c,d, from which the FSS and phase can be determined. Figure 2d shows a measurement close to the resolution limit of the setup given by a clear resolution of a FSS of ~ 2 μeV. It is worth mentioning that the average FSS values for our sample (FSS distribution between 2 to 35 μeV and average value of 10±6 μeV) are smaller than those of typical telecom-wavelength QDs reported in Ref.[40] (see Fig. 2e).

In order to prove the capability of strain tuning to eliminate the FSS, we investigated the neutral exciton emission lines of two different QDs under the application of uniaxial strain fields, see Fig. 3. The applied strain affects the emission properties in two ways. Firstly, we observe a linear tuning of the emission wavelengths as a function of the applied voltage $V_P$. The emission of the two QDs are shifted by 0.6 nm and 0.75 nm (Figs. 3 (a) and (d)), respectively. Secondly, as shown in Figs. 3 (b) and (e), the FSS values change from the initially low values to zero, i.e. below the resolution limit of the experimental setup. Similar to our previous observations, the elimination of the FSS is accompanied with a distinct phase shift of 90°,[24,39,41] The experimental data of the corresponding phase measurement on the two QDs are included in Figs. 3 (c) and (e), respectively.

The uniaxial strain tuning of the FSS has been investigated in several theoretical[17,18] and experimental[24,25] studies. It originates from the coherent coupling of the two bright excitonic states. The experimental data for the FSS and the phase presented in Fig. 3b-f is fitted to the theoretical model developed by M.Gong et al.[17] This calculation is based on the general relation of asymmetry and polarization angle and computes the FSS in QDs strain-tuned by means of piezoelectric supports. Under the condition of uniaxial stress, the QD's FSS with applied piezo voltage $V_P$ is computed via

$$\text{FSS}(V_P) = \sqrt{4(\beta V_P + \kappa)^2 + (\alpha V_P + 2\delta)^2} \tag{1}$$

where $\alpha$ and $\beta$ are related to the elastic compliance constants and the valence band deformation potentials[42]. For stress operating along the [110] and [1$\bar{1}$0] crystal directions $\beta = 0$.[17] The parameters $\kappa$ and $\delta$ account for the QD structural asymmetry and corresponding modulation of the exciton dipole moments.[42]

According to the same model, the polarization angle θ over $V_P$ can be obtained from:

$$\tan(\theta_\pm) = \frac{-2\delta - V_P \pm \text{FSS}(V_P)}{2(\beta V_P + \kappa)} = \frac{-2\delta - V_P \pm \sqrt{4(\beta V_P + \kappa)^2 + (\alpha V_P + 2\delta)^2}}{2(\beta V_P + \kappa)} \tag{2}$$

In addition to the change of the FSS we observe an abrupt change of 90° for the polarization angle in the spectrum while the FSS reaches its minimum, which is evidence of its elimination. Both QDs, as presented in Figure 3, show good agreement in their tuning behavior and clearly prove the reduction of the FSS below the present resolution limit. The parameters $\alpha$, $\delta$ and $\kappa$ fitted to the experimentally obtained FSS data for both QDs according to eq. 1 with $\beta = 0$ can be found in Table I.

TABLE I. Fitted parameters of eq. 1 for QD 1 and 2.

|  | QD 1 | QD 2 |
|---|---|---|
| α (μeV/V) | 0.11 ± 0.01 | 0.08 ± 0.01 |
| δ (μeV) | 3.64 ± 0.28 | -1.29 ± 0.2 |
| κ (μeV) | 0.36 ± 0.18 | 0.5 ± 0.38 |

From the aforementioned fit functions we can extract minimal accomplished FSS-values of 0.75 μeV and 1.01 μeV for QD 1 and 2, respectively. Due to the uniaxial strain tuning, the reduction of the FSS is limited to the alignment accuracy towards the QDs asymmetry axes, which is then again preferentially aligned along the crystal axes. Despite that, the achieved minimal FSS values are on the order of the lifetime-limited linewidth, thus enabling, in principle, high-efficiency generation of polarization-entangled photon pairs.

In conclusion, we have successfully demonstrated the capability to tune the QD emission energy and reduce the FSS for dots emitting at telecom wavelengths. These properties are of great interest for the realization of long distance quantum networks. On one hand, the possibility to deterministically control the source emission energy renders the efficient interference of several independent emitters possible. On the other hand, the reduction of the FSS is needed for the generation of polarization entangled photons, which are the basis for various entanglement distribution schemes. We have demonstrated that the intrinsically small FSS can be fully eliminated in MOVPE-grown QDs, for which the emission wavelength has been shifted to the telecom O-band.

As an outlook, the already demonstrated capability of incorporating a second, independent, tuning knob, *e.g.* via uniaxial strain in combination with Stark tuning,[42] would open the possibility of realizing a source of entangled photons at telecom wavelengths in which the energy tunability can be achieved independently, i.e. without varying the reduced FSS. In addition, as the suppression of FSS in semiconductor QDs via strain tuning was recently demonstrated for an on-chip platform,[6] a similar architecture for telecom wavelengths would be of great interest to implement a quantum repeater network.

ACKNOWLEDGMENTS: The work was financially supported by the BMBF Q.Com-H (16KIS0106) and Q.com-H (16KIS0115), and by the ERC Starting Grant "QD-NOMS" with No. 715770. The authors thank B. Eichler, R. Engelhard, and S. Harazim for the technical support in device fabrication.

REFERENCES:

[1] N. Akopian, N. H. Lindner, E. Poem, Y. Berlatzky, J. Avron, D. Gershoni, B. D. Gerardot, and P. M. Petroff, Physical Review Letters **96,** 130501 (2006).

[2] R. J. Young, R. M. Stevenson, P. Atkinson, K. Cooper, D. A. Ritchie, and A. J. Shields, New Journal of Physics **8,** 29 (2006).

[3] R. Hafenbrak, S. M. Ulrich, P. Michler, L. Wang, A. Rastelli, and O. G. Schmidt, New Journal of Physics **9,** 315 (2007).

4   A. Laucht, S. Putz, T. Gunthner, N. Hauke, R. Saive, S. Frederick, M. Bichler, M. C. Amann, A. W. Holleitner, M. Kaniber, and J. J. Finley, Physical Review X **2,** 7(2000).

5   G. Reithmaier, S. Lichtmannecker, T. Reichert, P. Hasch, K. Muller, M. Bichler, R. Gross, and J. J. Finley, Scientific Reports **3,** 6 (2013).

6   Y. Chen, J. X. Zhang, M. Zopf, K. B. Jung, Y. Zhang, R. Keil, F. Ding, and O. G. Schmidt, Nature Communications **7,** 7 (2016).

7   N. Somaschi, V. Giesz, L. De Santis, J. C. Loredo, M. P. Almeida, G. Hornecker, S. L. Portalupi, T. Grange, C. Anton, J. Demory, C. Gomez, I. Sagnes, N. D. Lanzillotti-Kimura, A. Lemaitre, A. Auffeves, A. G. White, L. Lanco, and P. Senellart, Nature Photonics **10,** 340 (2016).

8   Y. He, Y. M. He, Y. J. Wei, X. Jiang, M. C. Chen, F. L. Xiong, Y. Zhao, C. Schneider, M. Kamp, S. Hofling, C. Y. Lu, and J. W. Pan, Physical Review Letters **111,** 237404 (2013).

9   S. Unsleber, C. Schneider, S. Maier, Y. M. He, S. Gerhardt, C. Y. Lu, J. W. Pan, M. Kamp, and S. Hofling, Optics Express **23,** 32977 (2015).

10  P. Michler, *Quantum Dots for Quantum Information Technologies* (Springer, Berlin, 2017).

11  G. P. Agrawal, *Fiber-optic communication systems*, Vol. 4th ed. (Wiley, New York, 2010).

12  M. Paul, J. Kettler, K. Zeuner, C. Clausen, M. Jetter, and P. Michler, Applied Physics Letters **106,** 122105 (2015).

13  M. Paul, F. Olbrich, J. Hoschele, S. Schreier, J. Kettler, S. L. Portalupi, M. Jetter, and P. Michler, Applied Physics Letters **111,** 033102 (2017).

14  R. Al-Khuzheyri, A. C. Dada, J. Huwer, T. S. Santana, J. Skiba-Szymanska, M. Felle, M. B. Ward, R. M. Stevenson, I. Farrer, M. G. Tanner, R. H. Hadfield, D. A. Ritchie, A. J. Shields, and B. D. Gerardot, Applied Physics Letters **109,** 163104 (2016).

15  F. Olbrich, J. Höschele, M. Müller, J. Kettler, S. L. Portalupi, M. Paul, M. Jetter, and P. Michler, Applied Physics Letters **111,** 133106 (2017).

16  M. B. Ward, M. C. Dean, R. M. Stevenson, A. J. Bennett, D. J. P. Ellis, K. Cooper, I. Farrer, C. A. Nicoll, D. A. Ritchie, and A. J. Shields, Nature Communications **5,** 6 (2014).

17  M. Gong, W. W. Zhang, G. C. Guo, and L. X. He, Physical Review Letters **106,** 227401 (2011).

18  R. Singh and G. Bester, Physical Review Letters **104,** 196803 (2010).

19  R. J. Warburton, C. Schaflein, D. Haft, F. Bickel, A. Lorke, K. Karrai, J. M. Garcia, W. Schoenfeld, and P. M. Petroff, Nature **405,** 926 (2000).

20  M. Bayer, G. Ortner, O. Stern, A. Kuther, A. A. Gorbunov, A. Forchel, P. Hawrylak, S. Fafard, K. Hinzer, T. L. Reinecke, S. N. Walck, J. P. Reithmaier, F. Klopf, and F. Schafer, Physical Review B **65,** 195315 (2002).

21  R. Winik, D. Cogan, Y. Don, I. Schwartz, L. Gantz, E. R. Schmidgall, N. Livneh, R. Rapaport, E. Buks, and D. Gershoni, Physical Review B **95,** 235435 (2017).

22  A. Rastelli, F. Ding, J. D. Plumhof, S. Kumar, R. Trotta, C. Deneke, A. Malachias, P. Atkinson, E. Zallo, T. Zander, A. Herklotz, R. Singh, V. Krapek, J. R. Schroter, S. Kiravittaya, M. Benyoucef, R. Hafenbrak, K. D. Jons, D. J. Thurmer, D. Grimm, G. Bester, K. Dorr, P. Michler, and O. G. Schmidt, Physica Status Solidi B-Basic Solid State Physics **249,** 687 (2012).

| 23 | F. Ding, R. Singh, J. D. Plumhof, T. Zander, V. Krapek, Y. H. Chen, M. Benyoucef, V. Zwiller, K. Dorr, G. Bester, A. Rastelli, and O. G. Schmidt, Physical Review Letters **104**, 067405 (2010). |
|---|---|
| 24 | J. D. Plumhof, V. Krapek, F. Ding, K. D. Jons, R. Hafenbrak, P. Klenovsky, A. Herklotz, K. Dorr, P. Michler, A. Rastelli, and O. G. Schmidt, Physical Review B **83**, 1213023(R) (2011). |
| 25 | A. J. Bennett, M. A. Pooley, R. M. Stevenson, M. B. Ward, R. B. Patel, A. B. de la Giroday, N. Skold, I. Farrer, C. A. Nicoll, D. A. Ritchie, and A. J. Shields, Nature Physics **6**, 947 (2010). |
| 26 | B. D. Gerardot, S. Seidl, P. A. Dalgarno, R. J. Warburton, D. Granados, J. M. Garcia, K. Kowalik, O. Krebs, K. Karrai, A. Badolato, and P. M. Petroff, Applied Physics Letters **90**, 041101 (2007). |
| 27 | A. J. Hudson, R. M. Stevenson, A. J. Bennett, R. J. Young, C. A. Nicoll, P. Atkinson, K. Cooper, D. A. Ritchie, and A. J. Shields, Physical Review Letters **99**, 266802 (2007). |
| 28 | R. M. Stevenson, R. J. Young, P. See, D. G. Gevaux, K. Cooper, P. Atkinson, I. Farrer, D. A. Ritchie, and A. J. Shields, Physical Review B **73**, 033306 (2006). |
| 29 | M. Benyoucef, L. Wang, A. Rastelli, and O. G. Schmidt, Applied Physics Letters **95**, 261908 (2009). |
| 30 | A. Rastelli, A. Ulhaq, S. Kiravittaya, L. Wang, A. Zrenner, and O. G. Schmidt, Applied Physics Letters **90**, 073120 (2007). |
| 31 | L. Sapienza, R. N. E. Malein, C. E. Kuklewicz, P. E. Kremer, K. Srinivasan, A. Griffiths, E. Clarke, M. Gong, R. J. Warburton, and B. D. Gerardot, Physical Review B **88**, 155330 (2013). |
| 32 | R. B. Patel, A. J. Bennett, I. Farrer, C. A. Nicoll, D. A. Ritchie, and A. J. Shields, Nature Photonics **4**, 632 (2010). |
| 33 | S. L. Portalupi, M. Widmann, C. Nawrath, M. Jetter, P. Michler, J. Wrachtrup, and I. Gerhardt, Nature Communications **7**, 6 (2016). |
| 34 | A. Dousse, J. Suffczynski, R. Braive, A. Miard, A. Lemaitre, I. Sagnes, L. Lanco, J. Bloch, P. Voisin, and P. Senellart, Applied Physics Letters **94**, 121102 (2009). |
| 35 | E. Goldmann, M. Paul, F. F. Krause, K. Muller, J. Kettler, T. Mehrtens, A. Rosenauer, M. Jetter, P. Michler, and F. Jahnke, Applied Physics Letters **105**, 152102 (2014). |
| 36 | J. Kettler, M. Paul, F. Olbrich, K. Zeuner, M. Jetter, and P. Michler, Applied Physics B-Lasers and Optics **122**, 6 (2016). |
| 37 | R. Trotta, P. Atkinson, J. D. Plumhof, E. Zallo, R. O. Rezaev, S. Kumar, S. Baunack, J. R. Schroter, A. Rastelli, and O. G. Schmidt, Advanced Materials **24**, 2668 (2012). |
| 38 | B. Hofer, J. Zhang, J. Wildmann, E. Zallo, R. Trotta, F. Ding, A. Rastelli, and O. G. Schmidt, Applied Physics Letters **110**, 151102 (2017). |
| 39 | Y. Zhang, Y. Chen, M. Mietschke, L. Zhang, F. F. Yuan, S. Abel, R. Huhne, K. Nielsch, J. Fompeyrine, F. Ding, and O. G. Schmidt, Nano Letters **16**, 5785 (2016). |
| 40 | E. Goldmann, S. Barthel, M. Florian, K. Schuh, and F. Jahnke, Applied Physics Letters **103**, 212102 (2013). |
| 41 | J. X. Zhang, J. S. Wildmann, F. Ding, R. Trotta, Y. H. Huo, E. Zallo, D. Huber, A. Rastelli, and O. G. Schmidt, Nature Communications **7**, 1 (2015). |

[42] R. Trotta, E. Zallo, C. Ortix, P. Atkinson, J. D. Plumhof, J. van den Brink, A. Rastelli, and O. G. Schmidt, Physical Review Letters **109,** 147401 (2012).

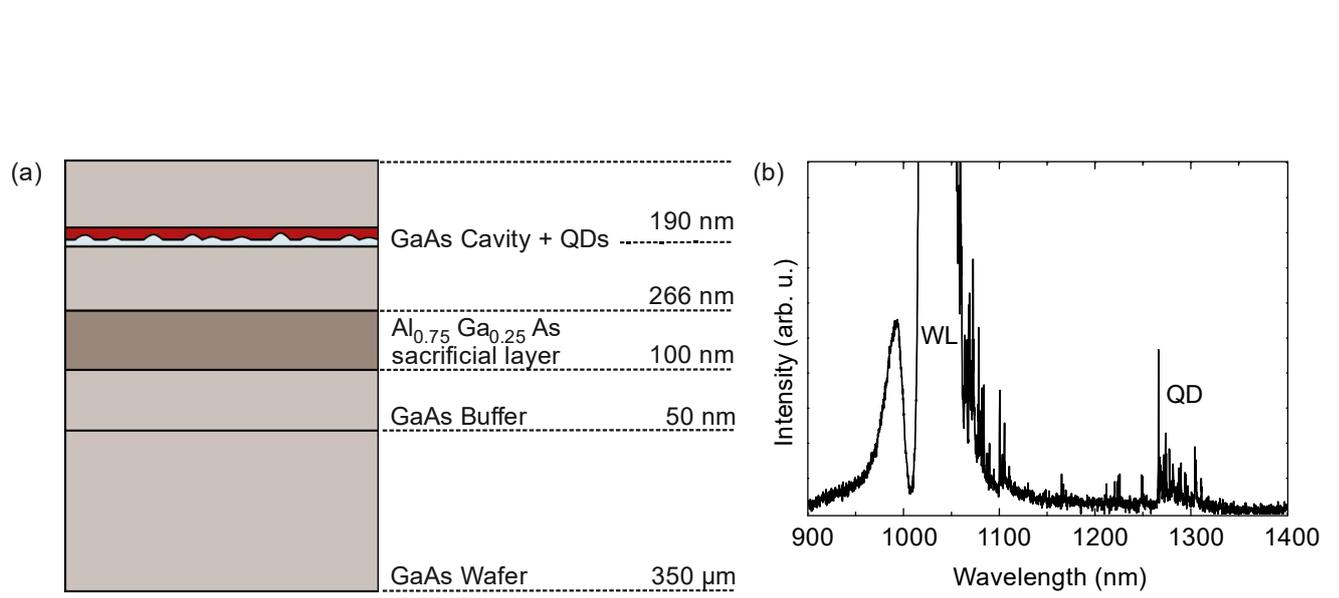

Figure 1: (a) Epitaxial layer structure of the grown sample. (b) Overview spectrum of the sample before transferring onto the piezo membrane. Several isolated single QD lines are observed inside the telecom O-band.

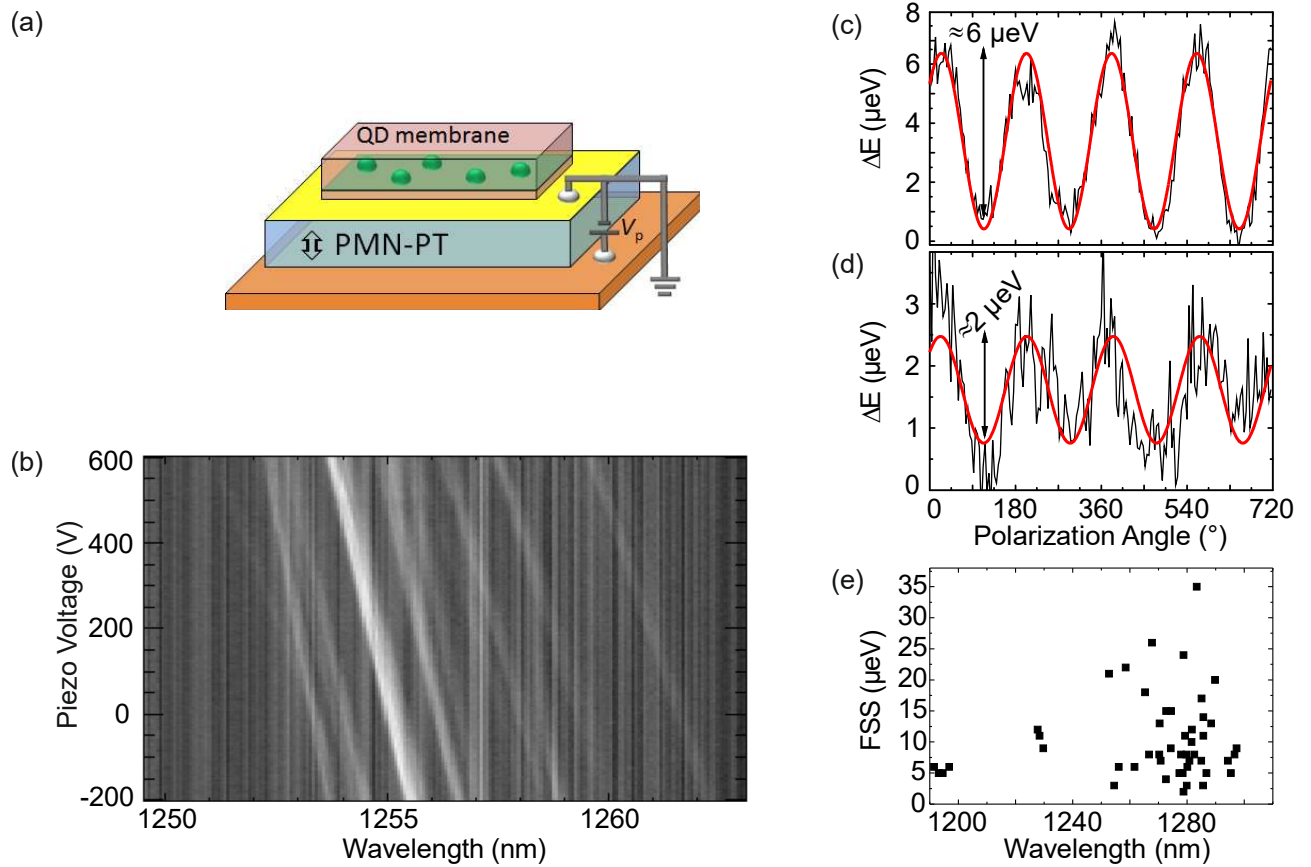

Figure 2: (a) Final device structure showing the QD-containing nanomembrane integrated onto a piezoelectric substrate. (b). Exemplary spectral tuning of a single QD. (c) Polarization series of selected transitions revealing a finite FSS, (d) second polarization series revealing the experimental resolution limit of about 2 µeV in the determination of the FSS (e) Statistics of FSS distribution for the as-grown sample.

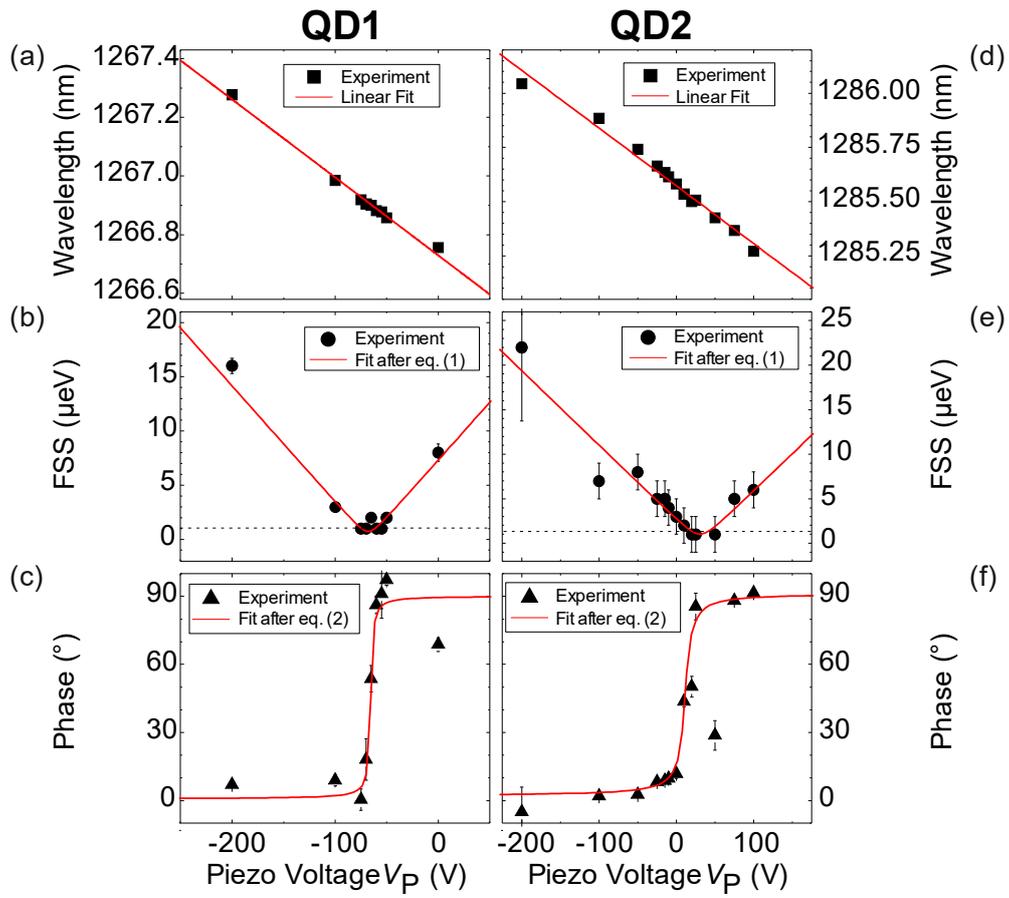

Figure 3: (a) & (d) Emission wavelength tuning over the applied voltage for QD1 and QD2 respectively. (b) & (e) FSS tuning under uniaxial strain employed to the QDs embedded in a nanomembrane structure (QD1 and QD2 respectively). (c) & (f) Voltage-dependent phase of one FSS component. The phase undergoes a characteristic jump of 90° when canceling the FSS (QD1 and QD2 respectively).